\documentclass[11pt,a4paper]{article} % JHEP
\pdfoutput=1

\usepackage{jheppub}                  % JHEP
\usepackage{here} 
\usepackage{graphicx}
\usepackage[T1]{fontenc}
\usepackage[normalem]{ulem}	% Part of the standard distribution

\def\lsim{\raise0.3ex\hbox{$\;<$\kern-0.75em\raise-1.1ex
\hbox{$\sim\;$}}}
\def\gsim{\raise0.3ex\hbox{$\;>$\kern-0.75em\raise-1.1ex
\hbox{$\sim\;$}}}

\title{On the robustness of IceCube's bound on sterile neutrinos in the presence of non-standard interactions}
\author{Arman Esmaili}
\author{and Hiroshi Nunokawa}
\affiliation{Departamento de F\'{\i}sica, Pontif{\'\i}cia Universidade Cat{\'o}lica do Rio de Janeiro, C. P. 38097, 22451-900, Rio de Janeiro, Brazil}
\emailAdd{arman@puc-rio.br}
\emailAdd{nunokawa@puc-rio.br}
\date{\today}
\abstract{The mixing parameters of sterile neutrino(s) preferred by the MiniBooNE and LNSD experiments are in strong tension with the exclusion limit from the IceCube experiment. Recently it has been claimed that by considering the non-standard neutrino interactions (NSI) in addition to the sterile neutrino, the IceCube's limit can be relaxed and the tension can be reconciled; a \textit{baroque} scenario as it has been called. We will show that this claim is just an artifact originating from the energy cuts of the chosen datasets. Contrary to the claim, by turning on the NSI and fixing the NSI parameters to the proposed values, not only the IceCube's limit on sterile neutrino cannot be alleviated, but in fact the tension will be aggravated (or at least keeps its strength). The reconciliation, more appropriately, can be called \textit{surreal}.}

\begin{document}

\maketitle

%%%%%%%%%%%%%%%%%%%%%%%%
%%%%%%%%%%%%%%%%%%%%%%%%
\section{Introduction\label{sec:introduction}}
%%%%%%%%%%%%%%%%%%%%%%%%
%%%%%%%%%%%%%%%%%%%%%%%%

Currently, almost all the neutrino data can be explained consistently in the $3\nu$ formalism, consisting of three active neutrino flavors and the corresponding mixing parameters (for a global fit to all the available data see~\cite{Esteban:2016qun,Capozzi:2018ubv,deSalas:2017kay}). However, there are still some \textit{anomalies}, coming from the electron neutrino appearance experiments, namely, the LSND~\cite{Aguilar:2001ty} and MiniBooNE~\cite{Aguilar-Arevalo:2013pmq} experiments, which indicate the existence of extra neutrino states, the so-called sterile neutrinos, with the mass $\sim\mathcal{O}(1)$~eV. In particular, the recent update from the MiniBooNE experiment~\cite{Aguilar-Arevalo:2018gpe}, which combines the $\nu_e$ and $\bar{\nu}_e$ appearance data, reports an excess of $4.8\sigma$ in the low energy range that can be increased to $6.1\sigma$ if combined with the LSND data. This excess can be interpreted in the $3+1$ scenario (3 active + 1 sterile neutrino state) with $\Delta m_{41}^2\sim\mathcal{O}(1)~{\rm eV}^2$ and $\sin^2 2\theta_{e\mu}\gtrsim 10^{-2}$. The allowed region in the $(\sin^2 2\theta_{e\mu},\Delta m_{41}^2)$ plane can be found in~\cite{Aguilar-Arevalo:2018gpe}.

The main obstacle in a happy interpretation of the LSND/MiniBooNE excess in terms of the sterile neutrinos is the strong tension with the disappearance data including MINOS/MINOS+~\cite{Adamson:2017uda} and IceCube~\cite{TheIceCube:2016oqi} experiments (for a global status of the sterile neutrino mixing from various experiments see~\cite{Dentler:2018sju,Gariazzo:2017fdh}). Among these the IceCube's limit has a different nature: while the MINOS experiment is sensitive to the sterile neutrino mixing through an averaged effect over the long baseline, the IceCube sensitivity originates from a resonance effect, an amplification of the $\bar{\nu}_\mu\to\bar{\nu}_s$ oscillation probability (or $\nu_\mu\to\nu_s$ for $\Delta m_{41}^2<0$) for atmospheric neutrinos crossing the Earth in the $\sim$~TeV energy range. The possibility of exploring the active-sterile neutrino mixing by looking at the energy and zenith angle distributions of the high energy atmospheric neutrinos has been proposed in~\cite{Yasuda:2000xs,Nunokawa:2003ep} (see also~\cite{Choubey:2007ji}). By the realization of the IceCube detector, as the first ${\rm km}^3$-volume neutrino telescope which is able to detect $\sim$~TeV atmospheric neutrinos, this possibility has been studied in detail: the limit on eV-scale sterile neutrinos has been derived using the data collected during the construction phase of the IceCube~\cite{Esmaili:2012nz} and it has been shown that few-years worth of the IceCube data can exclude completely the preferred region by LSND and MiniBooNE~\cite{Esmaili:2013vza}. The effect of the various mixing parameters in the $3+1$ scenario has been studied in~\cite{Esmaili:2013vza} and the search strategy generalized to the cascade event topology in~\cite{Esmaili:2013cja}. Finally the IceCube collaboration published the result of the sterile neutrino analysis~\cite{TheIceCube:2016oqi}, considering one-year muon-track data with the reconstructed muon energy proxy in the range $400~{\rm GeV}-20$~TeV, demonstrating the exclusion of the preferred parameter space region by the appearance experiments (to be precise, constraining the angle $\theta_{24}$ in the $3+1$ scenario).

Another new physics scenario that can be probed by the high energy atmospheric neutrinos observed by the IceCube is the non-standard neutrino interactions. This possibility has been proposed and used to derive the most stringent bound on the NSI parameters (the $\varepsilon_{\mu\tau}$ and $\varepsilon_{\mu\mu}-\varepsilon_{\tau\tau}$) in~\cite{Esmaili:2013fva}, with consistent results in~\cite{Salvado:2016uqu,Aartsen:2017xtt} (see~\cite{Esteban:2018ppq} for bounds on the NSI parameters from the global analysis of oscillation data). Recently it has been proposed in~\cite{Liao:2016reh} that the addition of non-standard neutrino interaction to the $3+1$ picture can relax the limit of IceCube on sterile neutrinos and reconcile the appearance and disappearance discrepant results. The same claim has been repeated in a more recent work~\cite{Liao:2018mbg} which analyses the data of MINOS+ and IceCube, including the DeepCore data, and concludes that a combination of the charged-current and neutral-current NSI can relax the limits of both experiments. In this framework, the charged-current NSI is required for the relaxation of the MINOS+ limit (a nonzero detector NSI, $\varepsilon_{\mu\mu}^D$, has been assumed) while the neutral-current NSI parameters (nonzero $\varepsilon_{\mu\mu}$, $\varepsilon_{\tau\tau}$ and $\varepsilon_{ss}$) will, pretendedly, loosen the IceCube's limits. The IceCube data analyzed in~\cite{Liao:2018mbg} consists of the publicly available sterile search data\footnote{\url{https://icecube.wisc.edu/science/data/IC86-sterile-neutrino}}~\cite{TheIceCube:2016oqi} with the muon energy\footnote{The public data is available in the muon energy range $[400~{\rm GeV},20~{\rm TeV}]$.} $\in [501~{\rm GeV},10~{\rm TeV}]$ and the DeepCore oscillation data\footnote{\url{https://icecube.wisc.edu/science/data/2018nuosc}}~\cite{Aartsen:2014yll} with the muon energy range $[6,56]~{\rm GeV}$.

In this paper we will argue that in fact the apparent relaxation of the IceCube's limit on the sterile neutrino in the $(3+1)+$NSI scenario originates from the negligence of the $[56,500]~{\rm GeV}$ energy range and has no physical rationale behind. It should be emphasized that although the recent IceCube publicization of data on atmospheric neutrinos does not include the $[56,500]~{\rm GeV}$ energy range, older data are available, such as the IC-79~\cite{Gross:2013iq} and IC-40~\cite{Abbasi:2010ie}, which cover this energy range and do not show any significant deviation from the $3\nu$ framework expectation. Considering this energy range, contrary to the claim in~\cite{Liao:2016reh,Liao:2018mbg}, the limit on sterile neutrino from IceCube will be stronger in the $(3+1)+$NSI scenario or will keep the strength, depending on the quality of the data in the $[56,500]~{\rm GeV}$ energy range. To this aim, actually, a simple oscillation probability calculation is enough to manifest the argument.

The paper is organized as follows: In sec.~\ref{sec:assumtion} we summarize the main features of the atmospheric muon (anti-)neutrino oscillation in the $(3+1)+$NSI scenario and describe our assumptions. In sec.~\ref{sec:prob} we discuss in detail, based mainly on oscillation probabilities, why the addition of NSI to the $3+1$ model cannot help to the reconciliation of the LSND/MiniBooNE and IceCube tension. Finally, in sec.~\ref{sec:conclusions}, we provide our conclusions.

%%%%%%%%%%%%%%%%%%%%%%
%%%%%%%%%%%%%%%%%%%%%%
\section{The $(3+1)+$NSI: a concert of new physics scenarios\label{sec:assumtion}}
%%%%%%%%%%%%%%%%%%%%%%
%%%%%%%%%%%%%%%%%%%%%%

The phenomenology of high energy atmospheric neutrino oscillation in the $3+1$ and NSI frameworks have been studied separately in~\cite{Esmaili:2013vza,Esmaili:2013fva}. The characteristics of oscillation in the $(3+1)+$NSI is just a simultaneous consideration of both frameworks. In the following we will summarize the evolution equation of neutrinos in the $(3+1)+$NSI scenario, in order to fix the notation and remind the main features.   

The evolution equation of the neutrinos, in the flavor basis, in the presence of matter effect, in the $(3+1)+$NSI scenario can be written as:
\begin{eqnarray}\label{eq:evolution}
i \frac{d}{dx} 
\left[
\begin{array}{c}
\nu_e \\
\nu_\mu \\
\nu_\tau \\
\nu_s 
\end{array}
\right]
= 
\left(H_0 + V_m(x)\right) 
\left[
\begin{array}{c}
\nu_e \\
\nu_\mu \\
\nu_\tau \\
\nu_s 
\end{array}
\right]~,
\end{eqnarray}
where $H_0$ corresponds to the Hamiltonian in vacuum, and is given by
\begin{eqnarray} 
H_0 = \frac{1}{2E_\nu}\; U_{3+1}
\left[
\begin{array}{cccc}
0 & 0 & 0 & 0  \\
0 & \Delta m^2_{21} & 0 & 0  \\
0 & 0 & \Delta m^2_{31} & 0  \\
0 & 0 & 0 & \Delta m^2_{41} 
\end{array}
\right]
U_{3+1}^{\dagger}~.
\end{eqnarray} 
Here $E_\nu$ is the neutrino energy, $\Delta m^2_{ij} \equiv m^2_i-m^2_j$ ($i,j = 1,2,3,4$) are the mass-squared differences, and $U_{3+1}$ is the mixing matrix for the $3+1$ model, parameterized as 
\begin{eqnarray}\label{eq:U}
U_{3+1} \equiv R_{34}(\theta_{34}) R_{24} (\theta_{24},\delta_{24}) R_{14}(\theta_{14}, \delta_{14}) R_{23}(\theta_{23}) R_{13} (\theta_{13}, \delta_{13}) R_{12}(\theta_{12})~,
\end{eqnarray} 
where $R_{ij}(\theta_{ij})$ is the rotation matrix with the angle $\theta_{ij}$ in the $i$-$j$ plane; while the rotation-like matrices $R_{ij}(\theta_{ij},\delta_{ij})$ can be obtained from $R_{ij}(\theta_{ij})$ by the following replacements: $\sin\theta_{ij}\to\sin\theta_{ij}e^{-i\delta_{ij}}$ and $-\sin\theta_{ij}\to-\sin\theta_{ij}e^{i\delta_{ij}}$. 

The $V_m$ term in Eq.~(\ref{eq:evolution}) represents the matter potential induced by the standard (coherent) interaction as well as by the NSI, which can be parameterized generally as
\begin{eqnarray}\label{eq:Hm}
V_m(x) = \sqrt{2}G_F n_e (x) 
\left[
\begin{array}{cccc}
1 + \varepsilon_{ee} & \varepsilon_{e\mu} & \ \varepsilon_{e\tau} &  \varepsilon_{es} \\
\varepsilon_{e\mu}^\ast & \varepsilon_{\mu\mu} & \ \varepsilon_{\mu\tau} &  \varepsilon_{\mu s} \\
\varepsilon_{e\tau}^\ast & \varepsilon_{\mu\tau}^\ast &
\  \varepsilon_{\tau\tau} &  \varepsilon_{\tau s} \\
\varepsilon_{es}^\ast & \varepsilon_{\mu s}^\ast & \ \varepsilon_{\tau s}^\ast &
 \kappa + \varepsilon_{ss}
\end{array}
\right]~, 
\end{eqnarray} 
where $G_F$ is the Fermi constant, $n_e$ is the electron number density of the propagation medium (in our case is the Earth), the $\varepsilon_{\alpha\beta}$ are the dimensionless parameters characterizing the strength of the NSIs\footnote{The Hermitian matrix of $\varepsilon$ parameters is effectively a sum over the NSI of neutrinos with the main ingredients of the Earth, that is the electron, $u$ and $d$ quarks. See~\cite{Esmaili:2013fva} for clarifications.}, and $\kappa \equiv n_n/(2n_e) \sim 1/2$ ($n_n$ being the neutron number density) for the Earth's matter. For anti-neutrinos, the overall sign of the matter potential in Eq.~(\ref{eq:evolution}) as well as that of the CP violating phases in Eq.~(\ref{eq:U}) must be flipped.

The oscillation pattern of the high energy atmospheric neutrinos in $3+1$ scenario is well-known: assuming $\Delta m_{41}^2>0$, where otherwise a strong tension with the cosmological data will appear, a resonance enhancement of the $\bar{\nu}_\mu\to\bar{\nu}_s$ oscillation occurs as neutrinos pass the Earth's core and/or mantle as follows. For core-crossing trajectories, that is $\cos\theta_z \lesssim -0.8$ where $\theta_z$ is the zenith angle, the enhancement is a parametric resonance~\cite{Liu:1997yb,Liu:1998nb} at the energy $\simeq (2.5~{\rm TeV})\cos2\theta_{24}~(\Delta m_{41}^2/{\rm eV}^2)$; while for the mantle-crossing trajectories ($\cos\theta_z \gtrsim -0.8$) the enhancement is an MSW resonance at the energy $\simeq(4~{\rm TeV})\cos2\theta_{24}~(\Delta m_{41}^2/{\rm eV}^2)$.   

The resonance enhancement of $\bar{\nu}_\mu\to\bar{\nu}_s$ conversion is independent of the $\theta_{14}$, and as shown in~\cite{Esmaili:2013vza} the enhancement becomes more effective with the increase of $\theta_{34}$. In what follows, we set $\sin^2\theta_{14}=0.02$ and $\theta_{34}=0$ where the latter choice leads to the most conservative limit on sterile neutrino mixing. For the standard $3\nu$-oscillation parameters we use the best-fit values of the global fit~\cite{Esteban:2016qun} for the normal mass ordering, $\Delta m^2_{31} > 0$, while our discussion will be equally valid also for the inverted mass ordering. It should be noticed that in the high energy range ($>100$~GeV) no standard oscillation effect is important and by decreasing the energy to $\sim20$~GeV the atmospheric oscillation parameters start to play a role. All the CP-violating phases also will be set to zero, as recommended in~\cite{Esmaili:2013vza} for a conservative limit.

In the presence of the NSI the resonance energies will be modified. To simplify the discussion, and also to consider the same setup as in~\cite{Liao:2016reh,Liao:2018mbg}, we will assume nonzero $(\varepsilon_{\mu\mu},\varepsilon_{\tau\tau},\varepsilon_{ss})$ and setting all the other NSI parameters to zero. By neglecting the $\nu_e$-flavor oscillation, see~\cite{Esmaili:2013vza}, through a simple calculation of the resonance condition in the $(3+1)+$NSI scenario we
can obtain the resonance energy of $\mu-s$ conversion. For the mantle-crossing trajectories the MSW resonance energy will be modified to 
\begin{equation}\label{eq:res}
E_\nu^{\rm res} \simeq 4~\cos 2 \theta_{24}
\left(\frac{\Delta m^2_{41}}{1\ \text{eV}^2} \right)
\left[\frac{1}{1-2\varepsilon_{\mu\mu}+2\varepsilon_{ss}} \right]~{\rm TeV}~.
\end{equation} 
For the core-crossing trajectories, where $\cos\theta_z\simeq-1$, the numerical factor of Eq.~({\ref{eq:res}}) should be replaced by $2.4$~TeV. As can be seen, the nonzero $\varepsilon_{\mu\mu}$ and/or $\varepsilon_{ss}$ shift the resonance energy. The resonance energy is independent of the $\varepsilon_{\tau\tau}$. Thus, neglecting for the moment the effect of the NSI parameters in the low energy, that is $E_\nu\sim(10-100)$~GeV, by turning on the neutral-current NSI the resonance will occur at some shifted energy, but does not disappear\footnote{Apparently, in Eq.~(\ref{eq:res}), it is possible to raise the resonance energy to a very high value by setting $\varepsilon_{\mu\mu}\simeq0$ and $\varepsilon_{ss}\simeq-1/2$, and thus, relax the IceCube bound. However, this setup of NSI parameters leads to anomalies in the cascade-type events of IceCube and can be constrained. We will leave this setup for a future study.}. The effect of the NSI parameters on the high energy atmospheric neutrino, including the low energy $(10-100)$~GeV range, has been studied in detail in~\cite{Esmaili:2013fva} and we will not repeat it here. The atmospheric oscillation probabilities in low energy range depend dramatically on the $\varepsilon_{\mu\mu}-\varepsilon_{\tau\tau}$. Thus, obviously, a large value inserted for $\varepsilon_{\mu\mu}$ in Eq.~(\ref{eq:res}) should be compensated with a large value for $\varepsilon_{\tau\tau}$ such that their difference remains small and the fit to the DeepCore data does not deteriorate. This is exactly the case chosen in~\cite{Liao:2016reh,Liao:2018mbg} and we will take it also in this paper.           

%%%%%%%%%%%%%%%%%%%%%
%%%%%%%%%%%%%%%%%%%%%
\section{Oscillation probabilities in the $(3+1)+$NSI scenario\label{sec:prob}}
%%%%%%%%%%%%%%%%%%%%%
%%%%%%%%%%%%%%%%%%%%%

The oscillation probabilities $P(\nu_\mu\to\nu_\mu)$ and $P(\bar{\nu}_\mu\to\bar{\nu}_\mu)$ can be obtained by the numerical solution of the neutrino evolution equation in the $(3+1)+$NSI scenario; {\it i.e.}, the Eq.~(\ref{eq:evolution}) and the corresponding one for the anti-neutrinos. In the numerical solution we will use the PREM model~\cite{Dziewonski:1981xy} for the Earth matter density profile assuming $Y_e = 0.5$, where $Y_e$ denotes the number of electrons per nucleon. 

By looking at the $(3+1)+$NSI oscillation probabilities, in this section we will argue and show that the relaxation of the IceCube's bound on the sterile neutrinos, as claimed in~\cite{Liao:2016reh,Liao:2018mbg}, does not happen. To be specific, we will consider two different sets of NSI parameters, motivated by and studied in~\cite{Liao:2018mbg}, which come from a scan over the NSI parameter values. These two sets, named case (a) and case (b), are summarized in Table.~\ref{tab:3+1-NSI_param}. For the case (a) the $\varepsilon_{ss}$ has been set to zero while a scan over $|\varepsilon_{\tau\tau}|<6$ and $|\varepsilon_{\tau\tau}-\varepsilon_{\mu\mu}|<0.5$ has been done\footnote{Although this set of values for the NSI parameters are in conflict with the solar neutrino data~\cite{Esteban:2018ppq}, we will continue with it as an example of the claimed maximum relaxation of IceCube's bound reported in~\cite{Liao:2018mbg}.}. For the case (b) the scan is over $|\varepsilon_{ss}|<6$, $|\varepsilon_{\tau\tau}|<0.5$ and $|\varepsilon_{\tau\tau}-\varepsilon_{\mu\mu}|<0.5$. As these two cases have been reported as the best scenarios of an scan over the NSI parameter values, any other choice of the NSI parameters
will be less effective in relaxing the IceCube's bound according to~\cite{Liao:2018mbg}.

The sterile neutrino mixing parameters in the cases (a) and (b) have been fixed to the reported best fitted values in~\cite{Liao:2018mbg}. Although, the $P(\nu_\mu\to\nu_\mu)$ and $P(\bar{\nu}_\mu\to\bar{\nu}_\mu)$ oscillation probabilities do not depend on $\theta_{14}$, we will just set it to $\sin^2\theta_{14}=0.02$ in our numerical calculation.       

%%%%%%%                   TABLE I                %%%%%%%%%%
%%%%%%%%%%%%%%%%%%%%%%%%%%%%%%%%
\begin{table}[h!]
\caption{The two sets of parameters in the $(3+1)+$NSI scenario considered in this work, which come from a scan of the parameter space performed in~\cite{Liao:2018mbg}. All the other parameters of the 3+1 model and/or NSI parameters not indicated in the table are assumed to be zero.\label{tab:3+1-NSI_param}}
\begin{center}%
\scalebox{1.0}{
\begin{tabular}{l|c|c|c|c|c|c} \hline \hline
Case & $\sin^2\theta_{14}$ &$\sin^2\theta_{24}$ & $\Delta m^2_{41}$ 
& $\epsilon_{\mu\mu}$ & $\epsilon_{\tau\tau}$ & $\epsilon_{ss}$
\\ \hline
(a) & 0.02 & 0.063  & 0.32 eV$^2$ & -4.3 & -4.0 & 0.0 \\
(b) & 0.02 & 0.032 & 0.62 eV$^2$   & -0.7 & -0.5 & 6.0 \\
\hline \hline
\end{tabular}%
}
\end{center}
\end{table}%
%%%%%%%%%%%%%%%%%%%%%%%%%%%%%%%%
%%%%%%%%%%%%%%%%%%%%%%%%%%%%%%%%

\begin{figure}[H]
%\vspace{-50mm}
\begin{center}
\vspace{-15mm}
\includegraphics[width=1.0\textwidth]{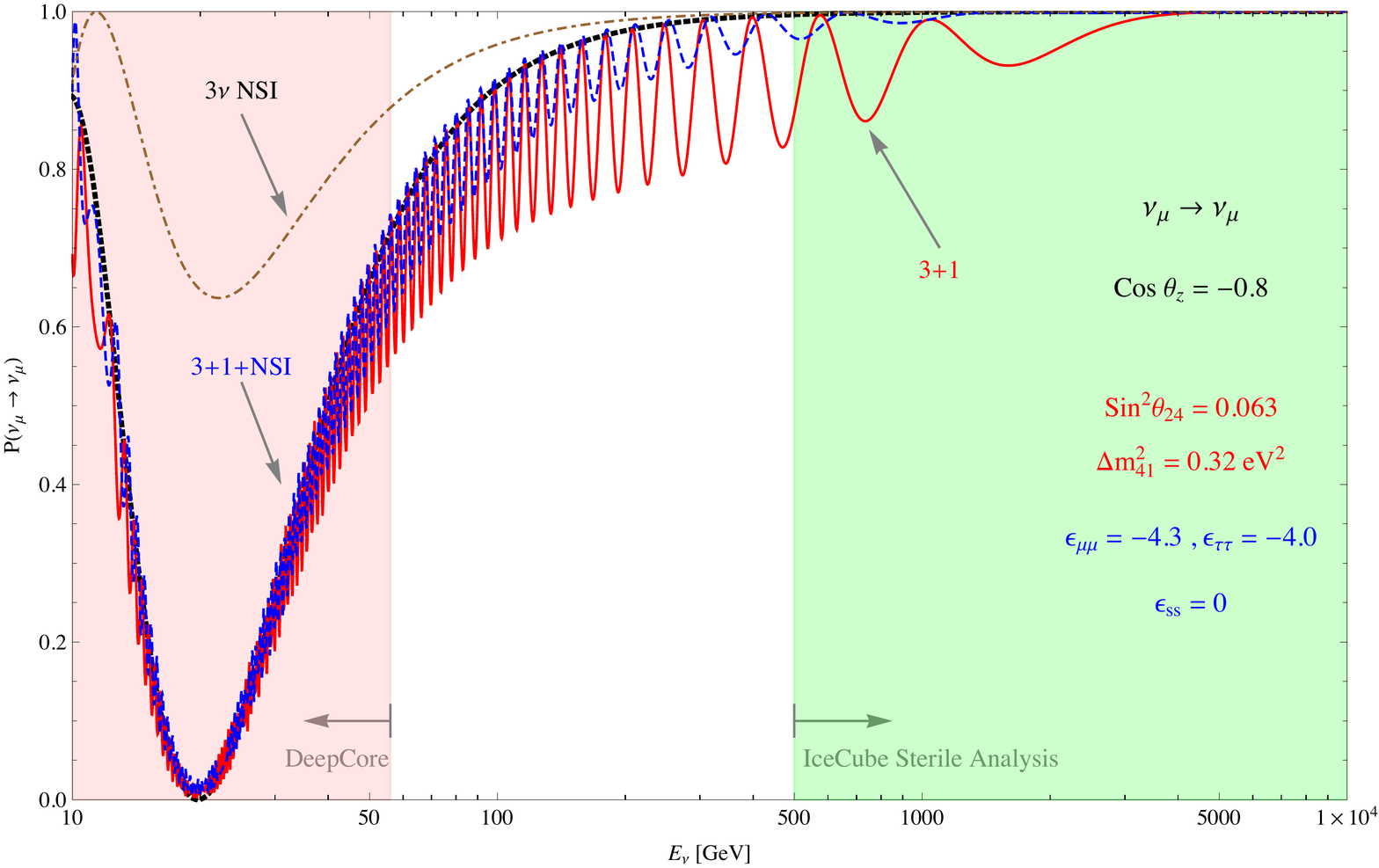}
\end{center}
\vspace{-14mm}
\begin{center}
\includegraphics[width=1.0\textwidth]{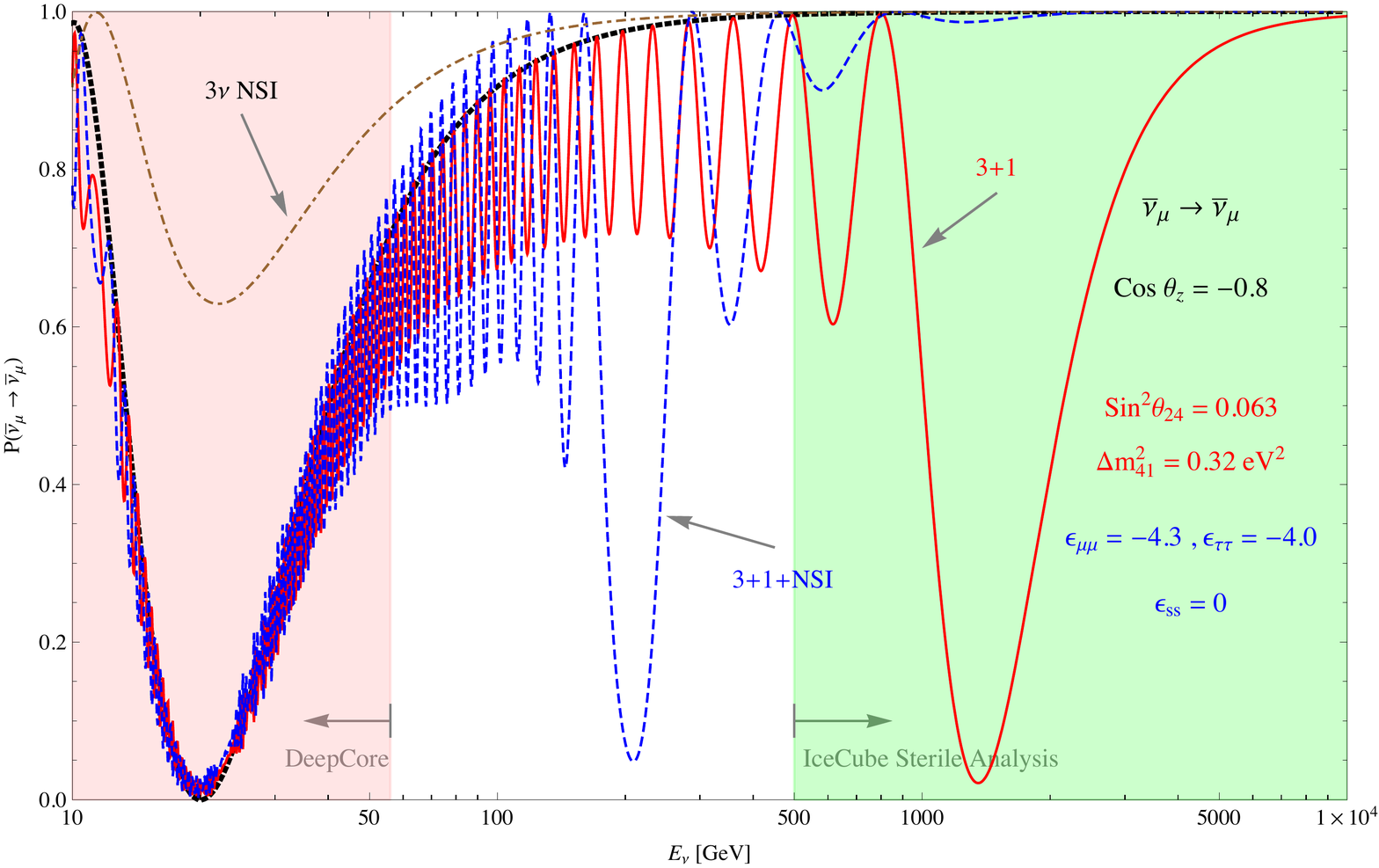}
\end{center}
\vspace{-5mm}
\caption{The $\nu_\mu \to \nu_\mu$ (upper panel) and $\bar{\nu}_\mu\to\bar{\nu}_\mu$ (lower panel) oscillation probabilities as a function of  the neutrino energy for $\cos \theta_z = -0.8$. The black thick dashed curve corresponds to the $3\nu$ oscillation, while the red solid curve corresponds to the $3+1$ model with $\sin^2\theta_{14} = 0.02$, $\sin^2\theta_{24} = 0.063$ and $\Delta m^2_{41}$ = 0.32 eV$^2$ (all the other parameters of the $3+1$ model are set to zero). The blue dashed curve indicates the case where the NSI is added on top of the $3+1$ model, the $(3+1)+$NSI scenario, with the parameters fixed to the case (a) shown in Table~\ref{tab:3+1-NSI_param}. For completeness, the case where only the NSI effect is added to the standard $3\nu$ oscillation is also shown by the brown dot-dashed curve. The energy ranges used by the IceCube's sterile neutrino analysis~\cite{TheIceCube:2016oqi} and the DeepCore oscillation analysis~\cite{Aartsen:2014yll} are indicated by the green and pink shaded regions, respectively.}
\label{fig:casea-z08}
\vspace{1mm}
\end{figure}

Let us start with the case (a). The upper and lower panels of the Figure~\ref{fig:casea-z08} show, respectively, the $P(\nu_\mu \to\nu_\mu)$ and $P(\bar{\nu}_\mu \to \bar{\nu}_\mu)$ oscillation probabilities for mantle-crossing trajectory $\cos\theta_z=-0.8$. The black thick dashed curve shows the oscillation probability in the $3\nu$ framework. The red solid curve shows the oscillation probability in the $3+1$ model with sterile mixing parameters as in case (a); {\it i.e.}, the mixing parameters shown in the first row of Table~\ref{tab:3+1-NSI_param} and setting the NSI parameters to zero. As expected, the $\Delta m_{41}^2=0.32~{\rm eV}^2$ leads to an MSW resonance at $E_\nu\simeq1.2$~TeV in the anti-neutrino channel. The IceCube's sensitivity to the mixing parameters of case (a) originates mainly from this dip in the anti-neutrino oscillation probability. By turning on the NSI parameters, the oscillation probability shown by blue dashed curve will be modified. The resonance dip in $\bar{\nu}_\mu\to\bar{\nu}_\mu$ oscillation, in accordance with Eq.~(\ref{eq:res}), shifts to $\sim200$~GeV. In Figure~\ref{fig:casea-z08} the green and pink shaded regions show, respectively, the energy ranges of the IceCube's sterile neutrino analysis~\cite{TheIceCube:2016oqi} and the DeepCore oscillation analysis~\cite{Aartsen:2014yll}. For comparison, the oscillation probability for $3\nu+$NSI scenario, with NSI parameter values of case (a), is also shown by the brown dot-dashed curve. As can be seen, addition of the sterile neutrino will makes the low energy part of the $3\nu+$NSI oscillation probability compatible with the $3\nu$, that otherwise is completely ruled out. Obviously, if one considers just the DeepCore and IceCube sterile neutrino analyses energy ranges, the limit on sterile neutrino in the $(3+1)+$NSI scenario can be relaxed; in the pink and green shaded regions there is a small difference between the $3\nu$ and $(3+1)+$NSI oscillation probabilities. However, the huge discrepancy between the $3\nu$ and $(3+1)+$NSI scenarios lies in the gap between the two energy ranges.

\begin{figure}[H]
%\vspace{-50mm}
\begin{center}
%\vspace{-5mm}
\includegraphics[width=1.0\textwidth]{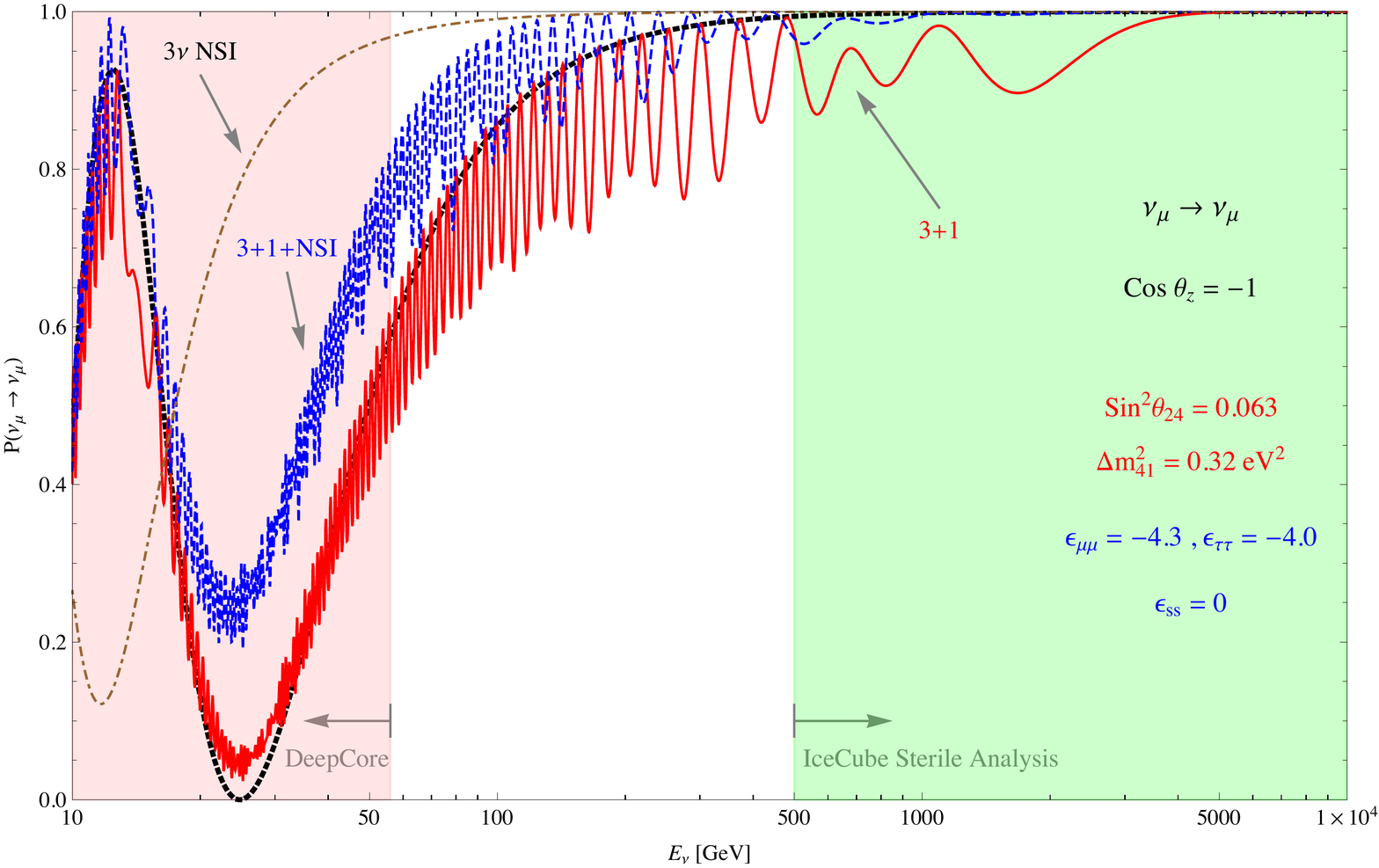}
\end{center}
\begin{center}
\includegraphics[width=1.0\textwidth]{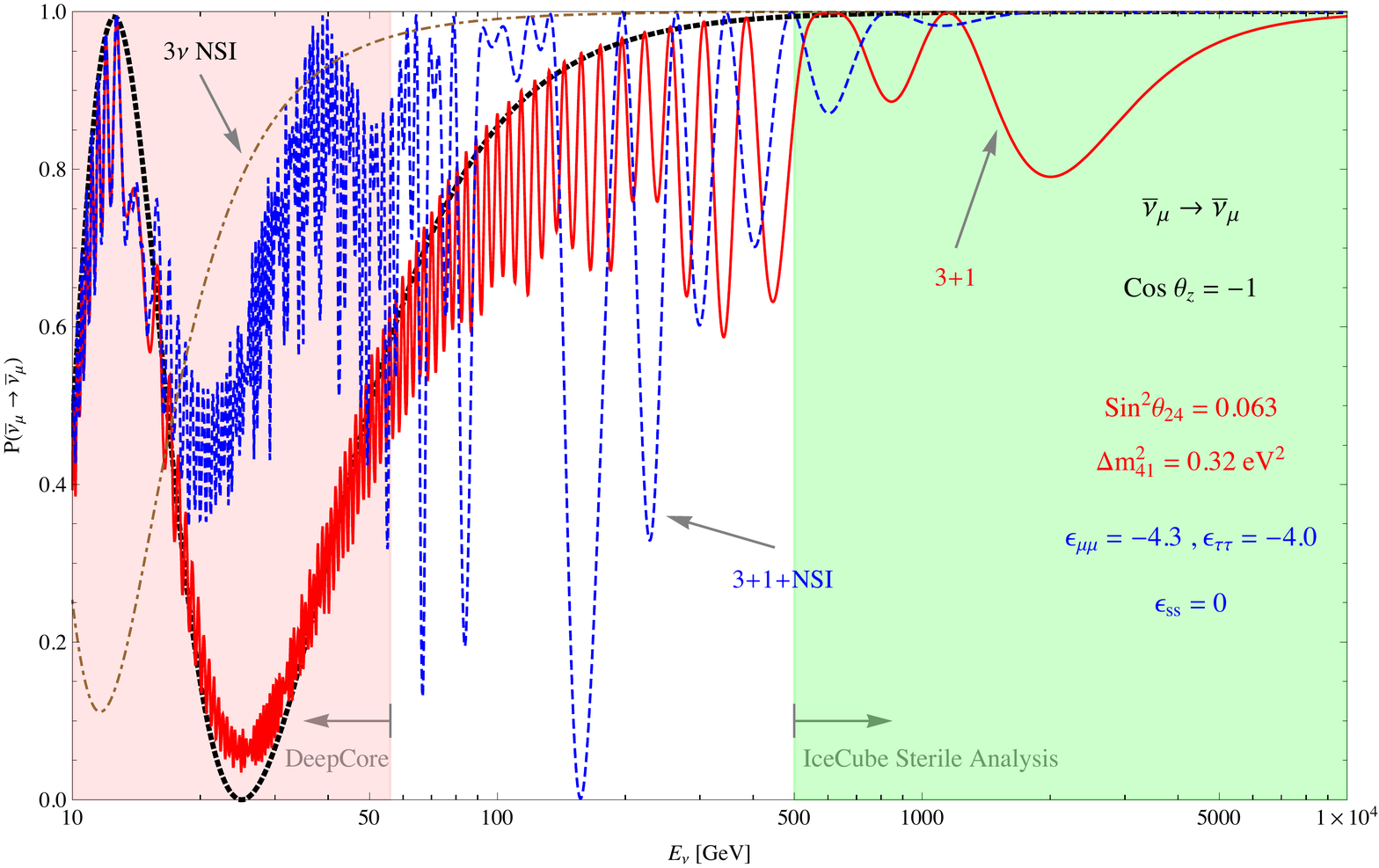}
\end{center}
\vspace{-5mm}
\caption{The same as Figure~\ref{fig:casea-z08} but for $\cos \theta_z = -1$.}
\label{fig:casea-z1}
\vspace{1mm}
\end{figure}

The Figure~\ref{fig:casea-z1} shows the oscillation probabilities for the case (a) and for the core-crossing trajectory with $\cos\theta_z=-1$, with the same color code and line type as in Figure~\ref{fig:casea-z08}. For $\cos\theta_z=-1$ the muon anti-neutrinos experience the parametric resonance in $3+1$ scenario at $\sim2$~TeV as expected; while in the $(3+1)+$NSI scenario the resonance shifts to lower energies, again in the gap between the energy ranges of DeepCore and IceCube sterile neutrino analyses. The are more deviations in the low energy part for the core-crossing trajectories due to the higher matter density in the propagation path of the neutrino which amplifies the effect of the NSI. In fact this deviation in the low energy part of the Figure~\ref{fig:casea-z1} is the origin of the limit obtained in~\cite{Liao:2018mbg} for case (a); otherwise, by just considering the green shaded region, IceCube sterile neutrino analysis is not sensitive to case (a).

\begin{figure}[H]
%\vspace{-50mm}
\begin{center}
\vspace{-1mm}
\includegraphics[width=1.0\textwidth]{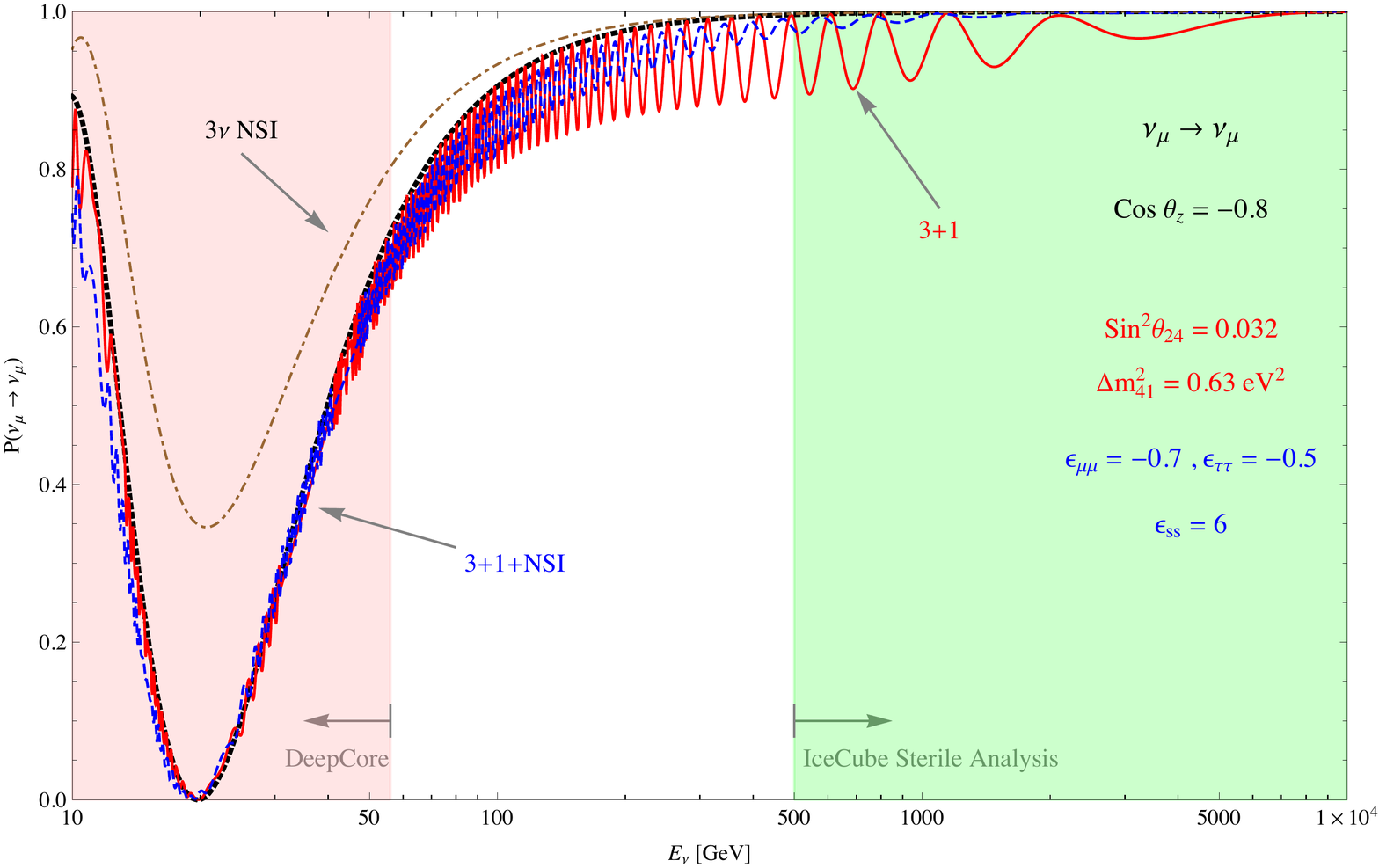}
\end{center}
\begin{center}
\includegraphics[width=1.0\textwidth]{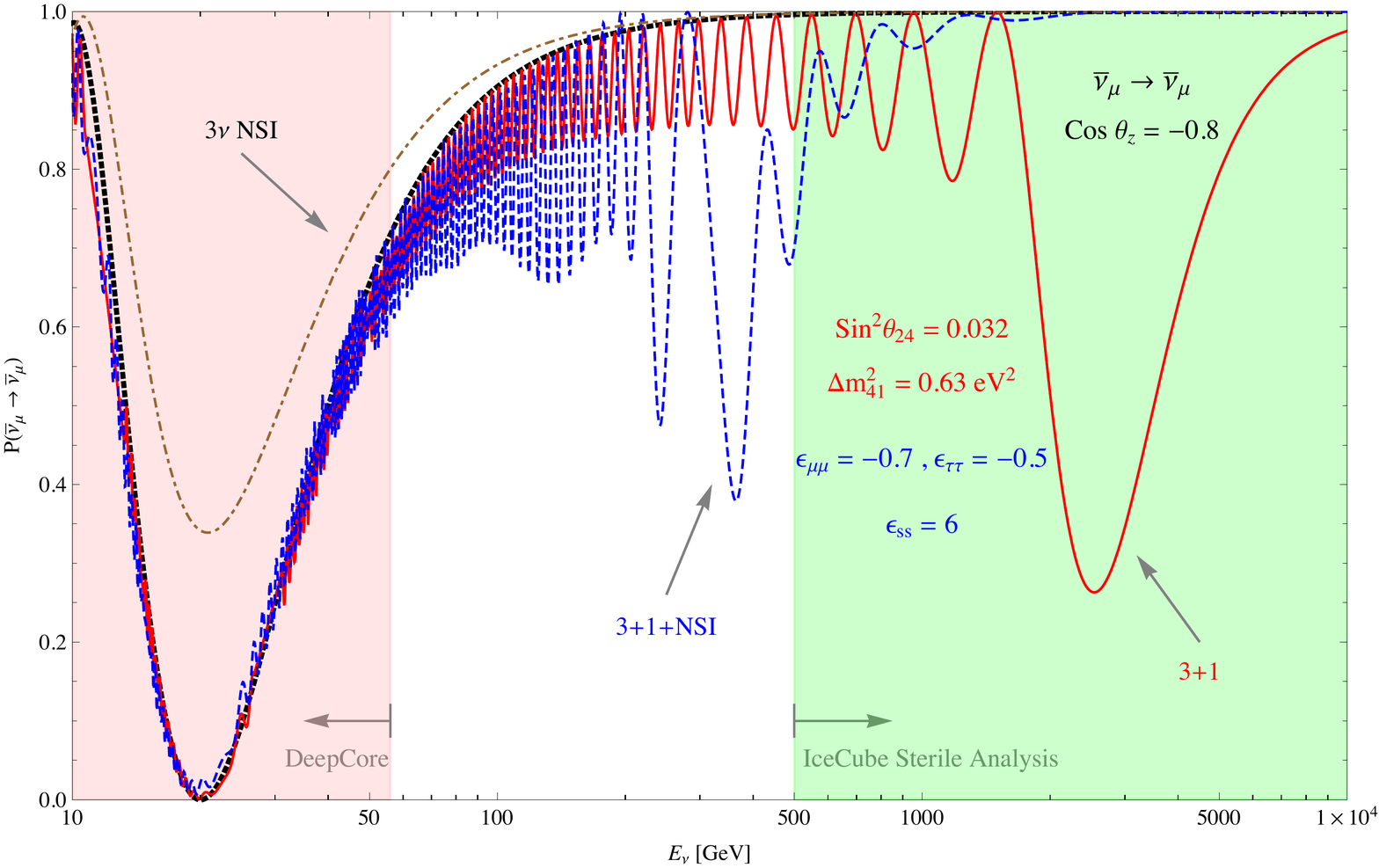}
\end{center}
\vspace{-5mm}
\caption{The same as Figure~\ref{fig:casea-z08} but for the case (b) shown in Table~\ref{tab:3+1-NSI_param}.}
\label{fig:caseb-z08}
\vspace{-5mm}
\end{figure}

\begin{figure}[H]
%\vspace{-50mm}
\begin{center}
\vspace{-5mm}
\includegraphics[width=1.0\textwidth]{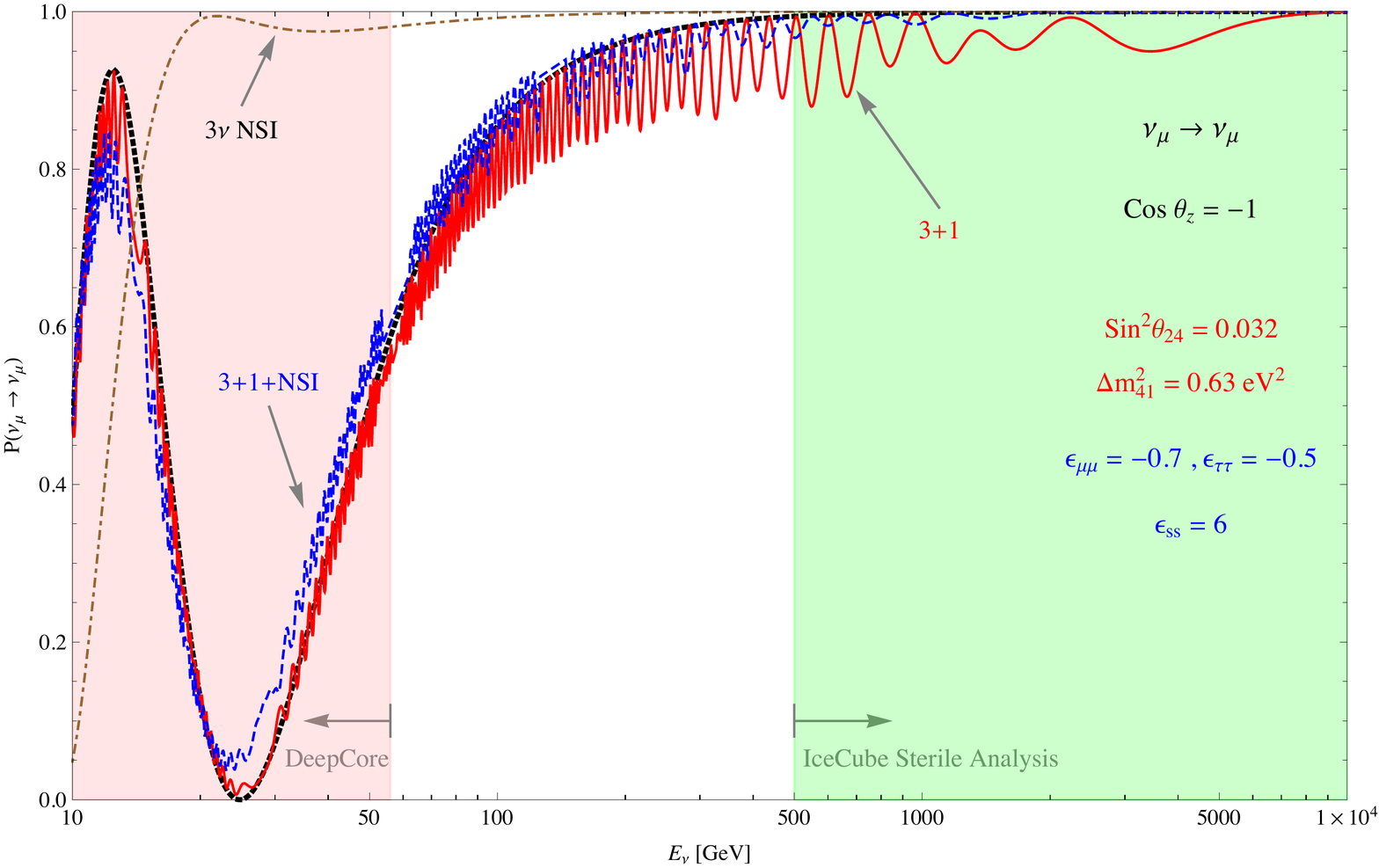}
\end{center}
\begin{center}
\includegraphics[width=1.0\textwidth]{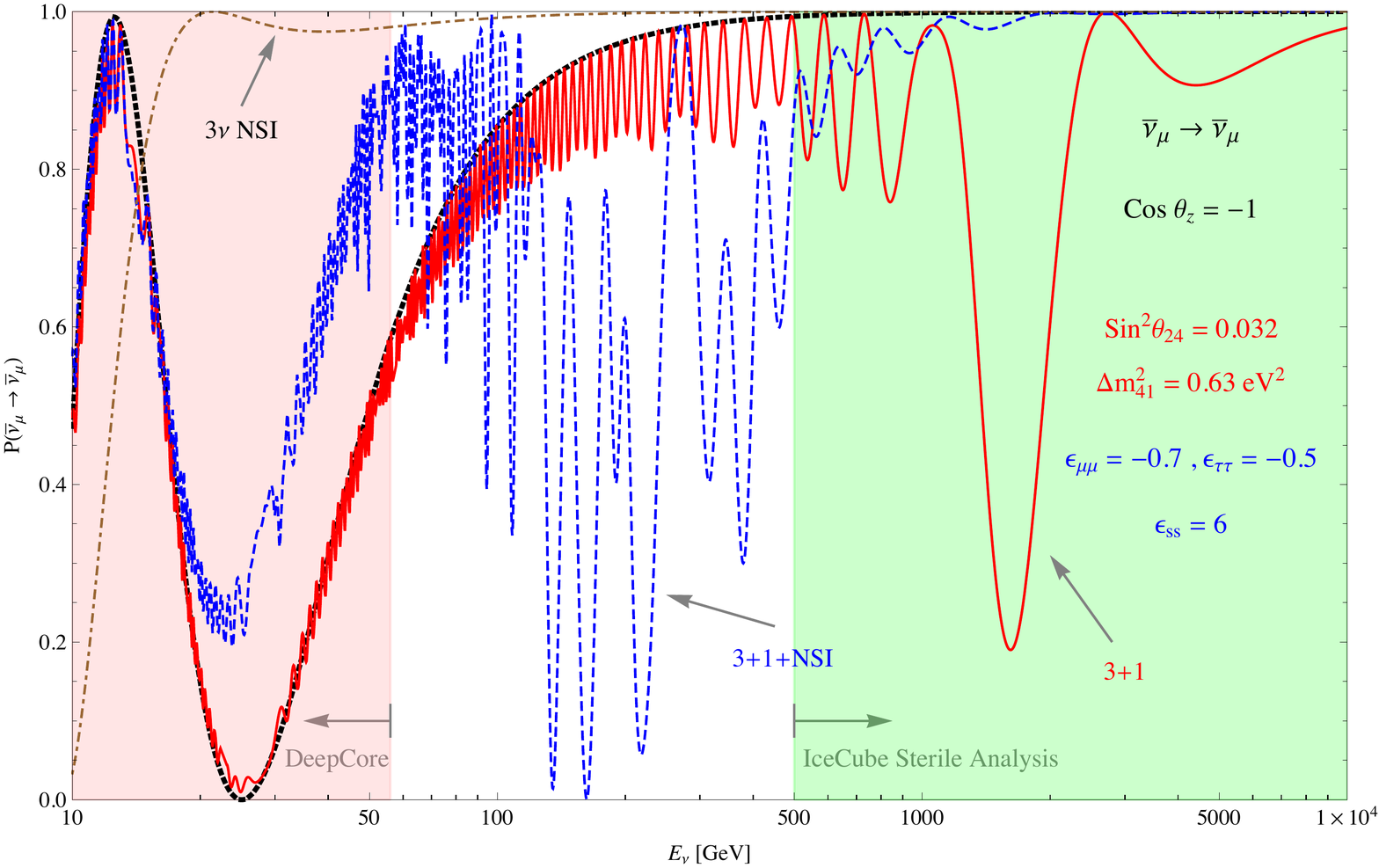}
\end{center}
\vspace{-5mm}
\caption{The same as in Fig.~\ref{fig:casea-z1} but for the case (b) shown in Table~\ref{tab:3+1-NSI_param}.}
\label{fig:caseb-z1}
\vspace{-5mm}
\end{figure}

Figures~\ref{fig:caseb-z08} and \ref{fig:caseb-z1} show the oscillation probabilities for the case (b), respectively, for the mantle and core crossing trajectories. The discussions presented for the case (a) apply also for the case (b). Again, the principal effect of the NSI is to shift the resonances into the gap in the energy ranges of the DeepCore and IceCube sterile neutrino analyses.

A comment on the energy ranges of the DeepCore and IceCube neutrino sterile analyses is in order: although the respective $[6,56]~{\rm GeV}$ and $[501~{\rm GeV},10~{\rm TeV}]$ energy ranges are referring to the muon energy produced in the $\nu_\mu$ and $\bar{\nu}_\mu$ charged-current interactions, we are taking it as a proxy of the neutrino energy and show them in the Figure~\ref{fig:casea-z08} (and the subsequent figures) as cuts on the $E_\nu$. Needless to say, this is just roughly correct and in a detailed analysis of the data the difference between the muon energy and $E_\nu$ should be taken into account. However, it is not our goal in this paper to perform an analysis of the data, basically since there are no publicly available data from IceCube collaboration in the energy range of $[56,500]~{\rm GeV}$. Instead, we will do a sensitivity analysis, similar to the one done in~\cite{Esmaili:2013vza}, just to quantify our argument. The main points of our argument are clear enough that hopefully motivate the IceCube collaboration to release the data in the whole energy range and perform an analysis by taking into account all the details.

Even without an elaborate calculation it is clear that a dip in the oscillation probability at $\sim300$~GeV can be excluded more strongly than or at least at the same level of a dip at $\sim2$~TeV; which is what happening in $(3+1)+$NSI scenario (see, for example, the lower panel of Figure~\ref{fig:casea-z08}). The atmospheric muon neutrino flux drops roughly as $\propto E_\nu^{-3.7}$ with the increase of energy. By taking into account the energy dependence of the cross section and the increase of effective volume by the increase in energy, the statistics is higher roughly by a factor of few at $E_\nu\sim500$~GeV with respect to $E_\nu\sim2$~TeV. For example, in the IC-79 dataset the peak of the statistics is at $\sim 500$~GeV, and the dataset contains as much events at $\sim300$~GeV as in $\sim2$~TeV (see figure~1 of~\cite{Gross:2013iq}). Other elements, such as the better energy resolution in the low energy due to the shorter muon range, also help toward a better constraining power. Thus, for sure, one can conclude that the $(3+1)+$NSI for both cases (a) and (b) is more strongly excluded than the $3+1$ model with the same sterile mixing parameters (or pessimistically it is excluded at the same level). This can be seen from the figures in this paper: by adding the NSI to the $3+1$ model not only the resonance dip in the muon neutrino survival oscillation probabilities slides to $\sim300$~GeV but also the oscillation probabilities will be modified at the DeepCore energy range. The sum of these two can lead to a strong bound from IceCube data. A detailed quantification of this statement is not possible for us since there is no recent public data from IceCube in the gap energy range. Thus, let us quantify a bit this statement by performing a sensitivity analysis using the information from the older IceCube datasets.  

For our sensitivity analysis we use the last public effective area of the IceCube in the whole range of energy, which goes back to the construction period of the detector: the IC-40~\cite{Abbasi:2010ie} and IC-79~\cite{Gross:2013iq} configurations. Of course, at least 10 times more data are available now and so we just increase the data-taking period correspondingly. For fixed values of the $(\sin^22\theta_{24},\Delta m_{41}^2)$ we calculate the $\chi^2$ value (which is the same as $\Delta\chi^2$ with respect to the $3\nu$ framework) by marginalizing the following $\chi^2$ function over the parameters $\alpha$ and $\beta$:  
\begin{equation}\label{eq:chi2,2}
\chi^2(\Delta m_{41}^2,\sin^2\theta_{24};\alpha,\beta)  = \sum_{i,j} \frac{\left\{N_{i,j}^0-\alpha [1+\beta(0.5+(\cos\theta_z)_i)]  N_{i,j}\right\}^2}{\sigma_{i,j,{\rm stat}}^2+\sigma_{i,j,{\rm sys}}^2} +  \frac{(1-\alpha)^2}{\sigma_\alpha^2} + \frac{\beta^2}{\sigma_\beta^2}~,
\end{equation}                  
where $\alpha$ and $\beta$ are the pull parameters taking into account the correlated uncertainties of the atmospheric neutrino flux normalization and its zenith dependence (tilt), respectively. Using the Honda flux of atmospheric neutrinos~\cite{Honda:2006qj}, these uncertainties are roughly $\sigma_\alpha=0.24$ and $\sigma_\beta=0.04$. The $N_{i,j}^0$ is the number of events in the $3\nu$ scenario in the $i^{\rm th}$ bin of energy and $j^{\rm th}$ bin of $\cos\theta_z$, which can be calculated by convoluting the effective area with the neutrino flux and oscillation probability. The $N_{i,j}$ is the number of events in the $(3+1)+$NSI scenario with the sterile mixing parameters $(\sin^2\theta_{24},\Delta m_{41}^2)$. The $\sigma_{i,j,{\rm stat}}=\sqrt{N_{i,j}}$ is the statistical error and we have added an uncorrelated systematic error $\sigma_{i,j,{\rm sys}}=fN_{i,j}$ where $f$ quantifies it.   

We have verified that by using the $[400~{\rm GeV},20~{\rm TeV}]$ energy range, 4 energy bins (uniformly distributed in log), 10 linearly distributed bins of $\cos\theta_z$ in $[-1,0]$, and an uncorrelated systematic error $f=10\%$ we can reproduce the exclusion plot of the IceCube analysis in~\cite{TheIceCube:2016oqi} almost exactly. Focusing on the case (b), the $\chi^2$ value for the $(\Delta m_{41}^2,\sin^2\theta)=(0.63~{\rm eV}^2,0.032)$ in the $3+1$ model using the $[500~{\rm GeV},10~{\rm TeV}]$ energy range is $\sim12$. By turning on the NSI parameters and going to $(3+1)+$NSI scenario, using the same energy range, the $\chi^2$ value drops to $\sim1$. This is in agreement with the claim in~\cite{Liao:2016reh,Liao:2018mbg} that the IceCube's limit on sterile neutrino relaxes in the presence of NSI. However, by extending the energy range to $[10~{\rm GeV},10~{\rm TeV}]$ the $\chi^2$ value increases to $\sim26$, as we expected. The same pattern, that is a significant increase of the $\chi^2$ value, occurs also for the case (a) if we extend the energy range as done for the case (b).

%%%%%%%%%%%%%%%%%%%%%%%%%
%%%%%%%%%%%%%%%%%%%%%%%%%
\section{Conclusions\label{sec:conclusions}}
%%%%%%%%%%%%%%%%%%%%%%%%%
%%%%%%%%%%%%%%%%%%%%%%%%%

The IceCube's bound on the active-sterile neutrino mixing(s) strongly excludes the parameter space preferred by the appearance experiments LSND and MiniBooNE, such that a global fit of the data in the $3+1$ scenario shows strong tensions. Recently it has been claimed that by adding non-standard neutrino interaction to the $3+1$ scenario it is possible to relax the IceCube's bound and weaken the tension~\cite{Liao:2016reh,Liao:2018mbg}.  

We have revisited the $(3+1)+$NSI scenario and studied the impact of the NSI on the IceCube's bound on the sterile neutrino. We have shown that in the presence of NSI the resonance enhancement of the $\bar{\nu}_\mu\to\bar{\nu}_s$ occurs at a shifted energy, but does not disappear. The reason behind the relaxation of the IceCube's bound claimed in~\cite{Liao:2016reh,Liao:2018mbg} is that for the chosen NSI parameters the resonance enhancement occurs in the gap $[56,500]~$GeV between the energy ranges of the two considered datasets (the IceCube sterile neutrino analysis~\cite{TheIceCube:2016oqi} and the DeepCore oscillation analysis~\cite{Aartsen:2014yll}). However, although the recent public data of the IceCube do not cover this gap, the older data such as IC-79~\cite{Gross:2013iq} and IC-40~\cite{Abbasi:2010ie} include this energy range and do not show any significant deviation from the $3\nu$ oscillation framework. By performing a sensitivity analysis, covering the energy range $[10~{\rm GeV},10~{\rm TeV}]$, we have shown that in fact the IceCube's bound on the sterile neutrino mixing becomes stronger in the presence of the NSI.

We therefore conclude that the NSI with the parameter values reported in~\cite{Liao:2018mbg} does not relax the IceCube's bound on the sterile neutrino. In the presence of NSI, the tension between the IceCube's limit and the LSND/MiniBooNE preferred region persists or even can get stronger. 

A detailed analysis by the IceCube collaboration is required to provide the correct limit on the $(3+1)+$NSI scenario. We hope that this work motivates the IceCube collaboration to publicize the atmospheric neutrino data in the full range of energy to avoid such misinterpretations.

%%%%%%%               FIGURE 1              %%%%%%%%%%%
%%%%%%%%%%%%%%%%%%%%%%%%%%%%%%%%
%%%%%%%%%%%%%%%%%%%%%%%%%%%%%%
%%%%%%%%%%%%%%%%%%%%%%%%%%%%%%

%%%%%%%               FIGURE 2              %%%%%%%%%%%
%%%%%%%%%%%%%%%%%%%%%%%%%%%%%%%%
%%%%%%%%%%%%%%%%%%%%%%%%%%%%%%%%
%%%%%%%%%%%%%%%%%%%%%%%%%%%%%%%%

%%%%%%%               FIGURE 3              %%%%%%%%%%%
%%%%%%%%%%%%%%%%%%%%%%%%%%%%%%%%
%%%%%%%%%%%%%%%%%%%%%%%%%%%%%%%%
%%%%%%%%%%%%%%%%%%%%%%%%%%%%%%%%

%%%%%%%               FIGURE 4              %%%%%%%%%%%
%%%%%%%%%%%%%%%%%%%%%%%%%%%%%%%%
%%%%%%%%%%%%%%%%%%%%%%%%%%%%%%%%
%%%%%%%%%%%%%%%%%%%%%%%%%%%%%%%%

\acknowledgments

A.~E. thanks the computing resource provided by CCJDR, of IFGW-UNICAMP with resources from FAPESP Multi-user Project 09/54213-0, and the partial support by the CNPq fellowship No.~310052/2016-5. H.~N. was supported by the CNPq research grants, No.~432848/2016-9 and No.~312424/2017-5.

\end{document}